\documentclass{ejm}

\usepackage{hyperref}
\usepackage{amsmath,commath,amsfonts,amssymb}
\usepackage{graphicx}
\usepackage{url}
\usepackage{bm}
\usepackage{natbib}

\pdfoutput=1

\ifprodtf \else
  \checkfont{eurm10}
  \iffontfound
    \IfFileExists{upmath.sty}
      {\typeout{^^JFound AMS Euler Roman fonts on the system,
                   using the 'upmath' package.^^J}%
       \usepackage{upmath}}
      {\typeout{^^JFound AMS Euler Roman fonts on the system, but you
                   don't seem to have the}%
       \typeout{'upmath' package installed. EJM.cls can take advantage
                 of these fonts,^^Jif you use 'upmath' package.^^J}%
       \providecommand\mathup[1]{##1}%
       \providecommand\mathbup[1]{##1}%
      }
  \else
    \providecommand\mathup[1]{#1}%
    \providecommand\mathbup[1]{#1}%
  \fi
\fi

\ifprodtf \else
  \checkfont{msam10}
  \iffontfound
    \IfFileExists{amssymb.sty}
      {\typeout{^^JFound AMS Symbol fonts on the system, using the
                'amssymb' package.^^J}%
       \usepackage{amssymb}%

      }{}
  \fi
\fi

\ifprodtf \else
  \IfFileExists{amsbsy.sty}
    {\typeout{^^JFound the 'amsbsy' package on the system, using it.^^J}%
     \usepackage{amsbsy}}
    {\providecommand\boldsymbol[1]{\mbox{\boldmath $##1$}}}
\fi

\DeclareSymbolFontAlphabet{\mathcal}   {symbols}
\DeclareSymbolFont{symbols}     {OMS}{cmsy}{m}{n}

\newcommand{\dat}[2]{\frac{ \partial #1}{ \partial #2 }}
\newcommand{\dpdS}[2]{\frac{\partial #1}{\partial #2}}
\newcommand{\hatz}{\hat{\bm z}}
\newcommand{\gradx}{\bm\nabla}
\newcommand{\gradxtwo}{\nabla^2}
\newcommand{\gradxfour}{\nabla^4}

\title[Linear Rayleigh-B\'{e}nard stability of a TI fluid]{Linear Rayleigh-B\'{e}nard stability of a transversely-isotropic fluid}

\author[Holloway, Smith \and Dyson]{%
  C.\ns R.\ns H\ls O\ls L\ls L\ls O\ls W\ls A\ls Y$\,^1$,\ns
  D.\ns J.\ns S\ls M\ls I\ls T\ls H$\,^{1,2}$\ns
\and
  R.\ns J.\ns D\ls Y\ls S\ls O\ls N$\,^{1*}$
}

\affiliation{%
  $^1\,$School of Mathematics, University of Birmingham, B15 2TT, U.K.\\
  $^2\,$Institute for Metabolism and Systems Research, University of Birmingham, B15 2TT, U.K.\\
   $^*\,$email\textup{\nocorr: \texttt{R.J.Dyson@bham.ac.uk}}}

\date{4 August 2017}
\pubyear{2000}
\volume{000}
\pagerange{\pageref{firstpage}--\pageref{lastpage}}

\begin{document}
\sloppy
\label{firstpage}
\maketitle

\begin{abstract}
Suspended fibres significantly alter fluid rheology, as exhibited in for example solutions of DNA, RNA and synthetic biological nanofibres. It is of interest to determine how this altered rheology affects flow stability. Motivated by the fact thermal gradients may occur in biomolecular analytic devices, and recent stability results, we examine the problem of Rayleigh-B\'{e}nard convection of the transversely-isotropic fluid of Ericksen.
A transversely-isotropic fluid treats these suspensions as a continuum with an evolving preferred direction, through a modified stress tensor incorporating four viscosity-like parameters. We consider the linear stability of a stationary, passive, transversely-isotropic fluid contained between two parallel boundaries, with the lower boundary at a higher temperature than the upper. 
To determine the marginal stability curves the Chebyshev collocation method is applied, and we consider a range of initially uniform preferred directions, from horizontal to vertical, and three orders of magnitude in the viscosity-like anisotropic parameters.
Determining the critical wave and Rayleigh numbers we find that transversely-isotropic effects delay the onset of instability; this effect is felt most strongly through the incorporation of the anisotropic shear viscosity, although the anisotropic extensional viscosity also contributes. Our analysis confirms the importance of anisotropic rheology in the setting of convection.
\end{abstract}


\section{Introduction}
Suspended fibres significantly alter the rheology of the fluid, as exhibited in for example suspensions of DNA \citep{marrington2005validation}, fibrous proteins of the cytoskeleton \citep{dafforn2004protein,kruse2005generic}, synthetic bio-nanofibres \citep{mclachlan2013calculations}, extracellular matrix \citep{dyson2015investigation} and plant cell walls \citep{dyson2010fibre}. It is of interest to determine how this altered rheology affects flow stability; motivated by the impact of anisotropic effects on Taylor-Couette instability \citep{holloway2015linear}, and the thermal gradients that may occur in devices which rely on nanofibre alignment for biomolecular analysis \citep{nordh1986flow}, we examine the Rayleigh-B\'{e}nard instability of the transversely-isotropic fluid of Ericksen.

\begin{figure}
\centering
\includegraphics[width=0.5\textwidth]{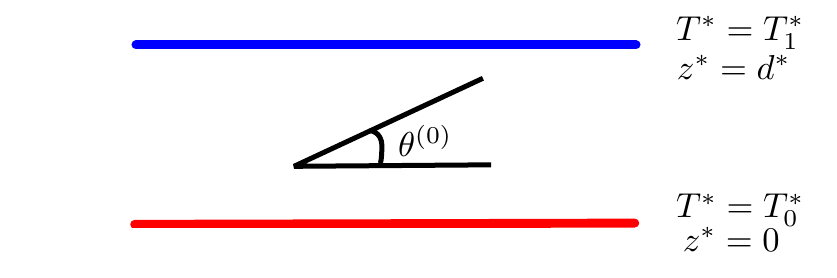}
\caption{A schematic diagram of the \citeauthor{rayleigh1916lix}-\citeauthor{benard} setup. The lower and upper boundaries are located at $z^*=0$ and $z^*=d^*$ at temperatures $T_0^*$ and $T_1^*$. The leading order preferred direction is given by the angle $\theta^{(0)}$. }\label{RBfig:Schematic}
\end{figure}

We consider the linear stability of a transversely-isotropic fluid contained between two infinitely-long horizontal boundaries of different temperatures (as shown in Figure \ref{RBfig:Schematic}), to a small arbitrary perturbation.  Three different combinations of boundary types will be considered, $(1)$ both boundaries are rigid, $(2)$ both are free, and $(3)$ the bottom boundary is rigid and the top is free. One application of our theory is to fibre-laden fluids, however it holds for any fluid which may be described as transversely-isotropic.

In this paper we adopt Ericksen's transversely-isotropic fluid \citep{ericksen1960transversely}, which has been used to describe fibre-reinforced media \citep{cupples2017viscous,dyson2015investigation,green2008extensional,holloway2015linear,lee2005continuum}. Ericksen's model consists of mass and momentum conservation equations together with an evolution equation for the fibre director field. The stress tensor depends on the fibre orientation and linearly on the rate of strain; it takes the simplest form that satisfies the required invariances. Recently Ericksen's model has been linked to suspensions of active particles \citep{holloway2017influences}, such as self-propelling bacteria, algae and sperm \citep{saintillan2013active}.

\citet{rayleigh1916lix} was the first to form a mathematical model of the Rayleigh-B\'{e}nard system, using equations for the energy and state of an infinite layer of fluid, bounded by two stationary horizontal boundaries of different constant uniform temperatures.
We work with the Boussinesq approximation that the flow is incompressible with non-constant density entering only through a buoyancy term.
Given we consider infinitesimal motion of a liquid the Boussinesq approximation is equally valid as for a Newtonian fluid.

In his original study \citet{rayleigh1916lix} was able to find a closed-form solution in the case of both upper and lower boundaries being free, \emph{i.e.}\ zero tangential stress; this setup has been simulated in experiments by replacing the bottom boundary with a layer of much less viscous fluid (leaving the top boundary free) \citep{Goldstein}. To determine the conditions where instability occurs for other combinations of boundary types, numerical techniques are required \citep{Drazin}.

We briefly discuss the equations and derive the steady state of the transversely-isotropic model (section \ref{RBsec:Governing_Equations}), and then undertake a linear stability analysis, leading to an eigenvalue problem which is solved numerically (sections \ref{RBsec:Stability}-\ref{RBsec:Numeric}).
The effect of variations in viscosity-like parameters and the steady state preferred direction on the marginal stability curves is considered (section \ref{RBsec:Results}), then we conclude with a discussion of the results in section \ref{RBsec:Discussion}.

\section{Governing equations}\label{RBsec:Governing_Equations}
We adopt a two dimensional Cartesian coordinate system $(x^*,z^*)$, and velocity vector ${\bm u}^*=(u^*,w^*)$; stars denote dimensional variables and parameters.
In formulating our governing equations we make use of the Boussinesq approximation \citep{chandrasekhar2013hydrodynamic}, treating the density as constant in all terms except bouyancy.
Mass conservation and momentum balance leads to the generalised Navier-Stokes equations
\begin{align}
\gradx^* \cdot \bm{u}^* &=\bm{0}, \label{RBeq:Mass_Conservation_Dimensional} \\
\rho_0^* \left( \dpdS{\bm{u}^*}{t^*} + \left( \bm{u}^* \cdot  \gradx^* \right) \bm{u}^* \right) &= - \gradx^* p^* + \gradx^* \cdot  \bm{\tau}^* - \rho^* g^* \hatz ,  \label{RBeq:Momentum_Dimensional}
\end{align}
where $\rho_0^*$ is the density at temperature $T_0^*$ of the lower boundary, $\rho^*(\bm{x}^*,t^*)$ is the variable density of the fluid, $t^*$ is time, $p^*$ is the pressure, $g^*$ is acceleration due to gravity, $\hatz$ is the unit vector in the $z^*$-direction, and $\bm{\tau}^*$ is the transversely-isotropic stress tensor proposed by \citet{ericksen1960transversely}, 
\begin{align}
\bm\tau^* &= 2 \mu^* \bm{e}^* + \mu_1^* \bm{a} \, \bm{a} + \mu_2^* \bm{a \, a\, a\, a : e}^* + 2 \mu_3^* \left( \bm{a \, a} \cdot \bm{e}^* + \bm{e}^* \cdot \bm{a \, a} \right) \label{RDimenStress}.
\end{align} 
Ericksen's stress tensor incorporates the single preferred direction $\bm{a}(\bm{x}^*,t^*)$, the rate-of-strain tensor $\bm{e}^* = (\bm\nabla^* \bm{u}^* + \bm\nabla^* \bm{u}^{*T})$ and viscosity-like parameters $\mu^*$, $\mu_1^*$, $\mu_2^*$ and $\mu_3^*$.
The parameter $\mu^*$ is the isotropic components of the solvent viscosity, modified by the volume fraction of the fibres \citep{dyson2010fibre,holloway2017influences}, $\mu_1^*$ implies the existence of a stress in the fluid even if it is instantaneously at rest, and can be interpreted as a tension in the fibre direction \citep{green2008extensional}, whilst the parameters $\mu_2^*$ and $\mu_3^*$ may be interpreted as the anisotropic extensional and shear viscosities respectively \citep{dyson2010fibre,green2008extensional,rogers1989squeezing}.

 We model the evolution of the fibre direction via the kinematic equation proposed by \citet{green2008extensional}
\begin{align}
\dpd{\bm{a}}{{t^*}} + \left( \bm{u}^* \cdot \bm\nabla^* \right) \bm{a} + \bm{a} \, \bm{a} \, \bm{a} : \bm\nabla^* \bm{u}^* = \left( \bm{a} \cdot \bm\nabla^* \right) \bm{u}^*, \label{TCeq:Kinematic_Condition}
\end{align}
which is a special case of the equation proposed by \citet{ericksen1960transversely}, appropriate for fibres with large aspect ratio. In the present study, we assume there is no active behaviour, \emph{i.e.}\ $\mu_1^*=0$ \citep{holloway2017influences}, therefore the stress tensor is given by
\begin{align}
\bm\tau^* &= 2 \mu^* \bm{e}^* + \mu_2^* \, \bm{a \, a\, a\, a : e}^* + 2 \mu_3^* \, \left( \bm{a \, a} \cdot \bm{e}^* + \bm{e}^* \cdot \bm{a \, a} \right) \label{RDimenStress}.
\end{align}

Temperature is governed by an advection-diffusion equation, 
\begin{align}
\dpdS{T^*}{t^*} + \left( \bm{u}^* \cdot \gradx^* \right) T^* &= \kappa^* \nabla^{*2} T^*,\label{RBeq:Temperature_Dimensional}
\end{align}
where $\kappa^*$ is the coefficient of thermal conductivity \citep{chandrasekhar2013hydrodynamic}, and the constitutive relation for density is given as
\begin{align}
\rho^* &= \rho_0^* \left( 1 - \alpha^* \left( T^* - T_0^* \right) \right), \label{RBeq:density_dimensional}
\end{align}
 which is a linear function of temperature and independent of pressure \cite{Drazin}. Here $\alpha^*$ is the coefficient of volume expansion and we have assumed both quantities $T^*$ and $\rho^*$ are independent of the fibres.
 
We will consider two types of bounding surfaces; for both types of surface we assume perfect conduction of heat and that the normal component of velocity is zero, \emph{i.e.}\
\begin{align}
\begin{split}
T^* = T_0^* & \text{ and } w^* = 0,  \text{ at }  z^* = 0, \\
T^* = T_1^* & \text{ and }  w^* = 0,  \text{ at }  z^* = d^*. 
\end{split}\label{RBeq:boundary_conditions1}
\end{align}
The distinction between the types of bounding surfaces is then made through the final two boundary conditions.
If the surface is rigid we impose no-slip boundary conditions, if the surface is free we impose zero-tangential stress, \emph{i.e.}\
\begin{align}
\begin{split}
u^* &= 0 \text{ on a rigid surface},\\
\dat{u^*}{x^*} &= 0 \text{ on a free surface}.
\end{split} \label{RBeq:boundary_conditions2}
\end{align}
Results will be presented from three groups of boundary conditions: both surfaces are rigid, both surfaces are free, and the bottom surface is rigid and the top surface is free.

\subsection{Non-dimensionalisation}

The model is non-dimensionalised by scaling the independent and dependent variables via:
\begin{align}
\bm{x}^* &= d^* \bm{x}, & t^* &= \frac{d^{*2}}{\kappa^*} t, & \bm{u}^* &= \frac{\kappa^*}{d^*} \bm{u} , \nonumber \\ T^* &= \beta^* d^* T, & (p^*, \bm\tau^*) &= \frac{\rho_0^* \kappa^{*2}}{d^{*2}} (p,\bm\tau), &\rho^* &= \rho_0^* \rho,
\end{align}
where variables without asterisks denote dimensionless quantities, and $\beta^*$ is the vertical temperature gradient, as chosen in \citet{Drazin}, \emph{i.e.}\ $\beta^* = (T_0^* - T_1^*)/d^*$.
The incompressibility condition \eqref{RBeq:Mass_Conservation_Dimensional} and the kinematic equation \eqref{TCeq:Kinematic_Condition} remain unchanged by this scaling,
\begin{align}
\gradx \cdot \bm{u} &=0, \label{RBeq:ND_mass} \\
 \dpd{\bm{a}}{t} + \left( \bm{u} \cdot \gradx \right) \bm{a} + \bm{a \, a \, a} : \gradx \bm{u} &= \left( \bm{a} \cdot \gradx \right) \bm{u}.
\label{RBeq:ND_Kinematic}
\end{align}
The momentum balance \eqref{RBeq:Momentum_Dimensional} becomes
\begin{align}
\dpd{\bm{u}}{t} + \left( \bm{u} \cdot \gradx \right) \bm{u} &= - \gradx p + \gradx \cdot \bm\tau - \frac{\mathcal{R} \mathcal{P}}{\mathcal{B}} \rho \hatz,	 \label{RBeq:ND_Momentum}
\end{align}
where we have introduced the following dimensionless parameters 
\begin{align}
\mathcal{B}&=\alpha^* \beta^* d^*,& \mathcal{R}&=\frac{\alpha^* \beta^* d^{*4} g^* \rho_0^* }{ \kappa \mu^*},&\mathcal{P} &= \frac{\mu^* }{\rho_0^* \kappa^*}.
\end{align} 
The Rayleigh number $\mathcal{R}$ is a dimensionless parameter relating the stabilising effects of molecular diffusion of momentum to the destabilising effects of buoyancy \citep{Drazin,koschmieder1993benard,sutton1950stability}, and the Prandtl number $\mathcal{P}$ relates the diffusion of momentum to diffusion of thermal energy \citep{chandrasekhar2013hydrodynamic}. 
Non-dimensionalising the stress tensor \eqref{RDimenStress} yields
\begin{align}
\bm\tau =& \mathcal{P} \left( 2 \, \bm{e} + \mu_2 \, \bm{a} \, \bm{a} \, \bm{a}\, \bm{a} : \bm{e} + 2 \mu_3 \left( \bm{a} \, \bm{a} \, \cdot \bm{e} + \bm{e} \cdot  \bm{a}\, \bm{a} \right) \right), \label{RBeq:ND_Stress}
\end{align}
where the non-dimensional rate-of-strain tensor, $\bm{e}=(\gradx \bm{u} + \gradx \bm{u}^T)/2$, and non-dimensional parameters
\begin{align}
 \mu_2 &= \frac{\mu_2^*}{\mu^*}, & \mu_3 &= \frac{\mu_3^*}{\mu^*},
\end{align}
have been introduced. 
Here $\mu_2$ and $\mu_3$ are the ratios of the extensional viscosity and shear viscosity in the fibre direction to the transverse shear viscosity, respectively \citep{green2008extensional,holloway2015linear}. 

The constitutive equation \eqref{RBeq:density_dimensional} for variable density is non-dimensionalised to give
\begin{align}
\rho = 1 + \mathcal{B} \left( T_0 - T \right) , \label{RBeq:ND_Density}
\end{align}
where $T_0=T_0^*/\beta^* d^*$ and equation \eqref{RBeq:Temperature_Dimensional}, which governs the temperature distribution, becomes
\begin{align}
\dpd{T}{t} + \left( \bm{u} \cdot \gradx \right)T &=\nabla^2 T. \label{RBeq:ND_Temperature}
\end{align}
Finally, the boundary conditions \eqref{RBeq:boundary_conditions1} and \eqref{RBeq:boundary_conditions2}, in dimensionless form, are
\begin{align}
\begin{split}
T = T_0, & \text{ and } w = 0,  \text{ at }  z = 0, \\
T = T_1, & \text{ and }  w = 0,  \text{ at }  z = 1, 
\end{split}\label{RBeq:ND_BC1}
\end{align}
where $T_1=T_1^*/\beta^* d^*$. The distinction between the type of surface remains unchanged,
\begin{align}
\begin{split}
u &= 0 \text{ on a rigid surface},\\
\dpd{u}{x} &= 0 \text{ on a free surface}.
\end{split} \label{RBeq:ND_BC2}
\end{align}

The model consists of four governing equations \eqref{RBeq:ND_mass}, \eqref{RBeq:ND_Kinematic}, \eqref{RBeq:ND_Momentum}, \eqref{RBeq:ND_Temperature} for $\bm{u},$ $\bm{a}$, $p$ and $T$, respectively, subject to constitutive laws \eqref{RBeq:ND_Stress} and \eqref{RBeq:ND_Density} with boundary conditions \eqref{RBeq:ND_BC1} and \eqref{RBeq:ND_BC2}.

\subsection{Steady state}\label{section:RBsteady}
Assuming that the parallel boundaries are infinitely long in the $x$-direction, a steady state solution is given by
\begin{align}
\bm{u}^{(0)} &= 0, & p^{(0)} &= p_0 - \frac{ \mathcal{RP}}{\mathcal{B}} \left( z + \frac{ \mathcal{B} \, z^2}{2} \right), \nonumber\\ T^{(0)} &=T_0- z, & \rho^{(0)} &= 1 + \mathcal{B} \,  z, \nonumber \\
\theta^{(0)} &= \text{constant}, & \bm{a}^{(0)}  &= (\cos \theta^{(0)}, \sin \theta^{(0)}), \label{rbsteady3}
\end{align}
where $p_0$ is some arbitrary pressure constant and the preferred fibre direction is described by the constant angle $\theta^{(0)}$ to the $x$-axis (Figure \ref{RBfig:Schematic}).

\section{Stability}\label{RBsec:Stability}
We now examine the linear stability of the steady state described by equations \eqref{rbsteady3}, for the three different combinations of boundary types. We derive the first-order equations for an arbitrary perturbation, which are transformed into a generalised eigenvalue problem by assuming the solution takes the form of normal modes. 

\subsection{Linear stability analysis}
We consider the stability of the steady state solution to a perturbation,
\begin{align}
\bm{u} (x,z,t) &=\hspace{1.1cm} \varepsilon \bm{u}^{(1)}(x,z,t) + \mathcal{O} \left( \varepsilon^2 \right),  \label{427}\\
 p(r,z,t) &= p^{(0)} + \varepsilon p^{(1)}(x,z,t) + \mathcal{O} \left( \varepsilon^2 \right), \\
  T(x,z,t) &= T^{(0)} + \varepsilon T^{(1)}(x,z,t) + \mathcal{O} \left( \varepsilon^2 \right), \\
 \theta(x,z,t) &= \theta^{(0)} + \varepsilon \theta^{(1)}(x,z,t) + \mathcal{O} \left( \varepsilon^2 \right). \label{RBeq:theta}
\end{align}
where $0 < \varepsilon \ll 1$. 
As we have proposed a perturbation to the fibre orientation angle $\theta^{(0)}$, and not the alignment vector $\bm{a}$ directly, the form of $\bm{a}$ is given by \citep{cupples2017viscous}
\begin{align}
\bm{a} &= \left( \cos \theta^{(0)} - \varepsilon \theta^{(1)} \sin \theta^{(0)}, \sin \theta^{(0)} + \varepsilon \theta^{(1)} \cos \theta^{(0)} \right) + \mathcal{O} \left( \varepsilon^2 \right). \label{436}
\end{align}
Here we have utilised the Taylor expansions for $\cos \theta$ and $\sin \theta$.

Using the ansatz given in equations \eqref{427}-\eqref{436} we may state the following governing equations at first order. The incompressibility condition \eqref{RBeq:ND_mass} becomes
\begin{align}
\gradx \cdot \bm{u}^{(1)} &= 0,	 \label{rb:firstMass}
\end{align}
with conservation of momentum \eqref{RBeq:ND_Momentum} given by
\begin{align}
	\dpd{\bm{u}^{(1)}}{t} &= - \gradx p^{(1)} + \gradx \cdot \bm{\tau}^{(1)} - \frac{ \mathcal{RP}}{\mathcal{B}} \rho^{(1)} \hatz.	 \label{rb:firstMomentum}
\end{align}
The first order constitutive relations for stress \eqref{RBeq:ND_Stress} and fluid density \eqref{RBeq:ND_Density} are given by
\begin{align}
\bm{\tau}^{(1)} &= \mathcal{P} \left( 2 \bm{e}^{(1)} + \mu_2 \bm{a}^{(0)} \bm{a}^{(0)} \bm{a}^{(0)} \bm{a}^{(0)} : \bm{e}^{(1)} + 2 \mu_3 \left( \bm{a}^{(0)} \bm{a}^{(0)} \cdot \bm{e}^{(1)} + \bm{e}^{(1)} \cdot \bm{a}^{(0)} \bm{a}^{(0)} \right) \right),  \label{rb:firstStress}\\
\rho^{(1)} &= - \mathcal{B}T^{(1)},	 \label{rb:firstDensity}
\end{align}
where $\bm{e}^{(1)} =( \gradx \bm{u}^{(1)} + (\gradx \bm{u}^{(1)})^T)/2$ is the first order rate-of-strain tensor.
Notice that equations \eqref{rb:firstMass}-\eqref{rb:firstDensity} are independent of the first order alignment vector 
\begin{align}
\bm{a}^{(1)}=\left( - \theta^{(1)} \sin \theta^{(0)}, \theta^{(1)} \cos \theta^{(0)} \right),	
\end{align}
 which is in turn governed by 
\begin{align}
	\dpd{\bm{a}^{(1)}}{t} + \bm{a}^{(0)} \bm{a}^{(0)} \bm{a}^{(0)} : \gradx \bm{u}^{(1)} &= \left( \bm{a}^{(0)} \cdot \gradx \right) \bm{u}^{(1)}. \label{rbkinematic}
\end{align}
Finally, the equation governing temperature at next order is
\begin{align}
\dpd{T^{(1)}}{t} - w^{(1)} &=\gradxtwo T^{(1)}. \label{RBeq:L2}
\end{align}
The boundary conditions become homogeneous at first order, and are given by
\begin{align}
w^{(1)} = u^{(1)} &= T^{(1)} = 0 \text{ on a rigid surface}, \\
w^{(1)} = \dpd{u^{(1)}}{x} &= T^{(1)} =0 \text{ on a free surface}.	
\end{align}
After eliminating pressure and substituting for stress, the components of the momentum equation \eqref{rb:firstMomentum} are given by\begin{align}
\begin{split}
\frac{1}{\mathcal{P}}  \gradxtwo \left(  \dpd{u^{(1)}}{t} \right)=& \left( 1 + \mu_2 \frac{\sin^2 2\theta^{(0)}}{4} + \mu_3 \right)  \gradxfour u^{(1)} \\ &+ \mu_2 \left( \frac{\sin 4 \theta^{(0)}}{2} \left( \dmd{u^{(1)}}{4}{x}{}{z}{3} - \dmd{u^{(1)}}{4}{x}{3}{z}{} \right) + \cos 4 \theta^{(0)} \dmd{u^{(1)}}{4}{x}{2}{z}{2} \right), 
\end{split} \label{RBeq:L4} \\
\begin{split}
\frac{1}{\mathcal{P}}  \gradxtwo \left( \dpd{w^{(1)}}{t} \right) =& \left( 1 + \mu_2 \frac{\sin^2 2\theta^{(0)}}{4} + \mu_3 \right) \gradxfour w^{(1)} + \mathcal{R} \dpd[2]{T^{(1)}}{x}  \label{RBeq:L1}  \\ &+ \mu_2 \left( \frac{\sin 4 \theta^{(0)}}{2} \left( \dmd{w^{(1)}}{4}{x}{}{z}{3} - \dmd{w^{(1)}}{4}{x}{3}{z}{} \right) + \cos 4 \theta^{(0)} \dmd{w^{(1)}}{4}{x}{2}{z}{2} \right).
\end{split}
\end{align}

Manipulating the components of the kinematic equation \eqref{RBeq:ND_Kinematic} yields an equation for the evolution of fibre direction,
\begin{align}
\dpd{\theta^{(1)}}{t} &= \cos^2 \theta^{(0)} \dpd{w^{(1)}}{x} - \sin^2 \theta^{(0)} \dpd{u^{(1)}}{z} - \sin 2 \theta^{(0)} \dpd{u^{(1)}}{x},  \label{RBeq:L3}
\end{align}
Notice equations \eqref{RBeq:L4}, \eqref{RBeq:L1} and \eqref{RBeq:L3} are decoupled, and so we may solve the stability problem by considering only equations \eqref{RBeq:L2} and \eqref{RBeq:L1} with appropriate boundary conditions on $w^{(1)}$ and $T^{(1)}$. The $x$-component of velocity and alignment angle may then be calculated from the solution for $w^{(1)}$.

We propose the solution to equations \eqref{RBeq:L2} and \eqref{RBeq:L1} takes the form
\begin{align}
w^{(1)} &= w'(z) e^{st+ikx}, & T^{(1)} &= T'(z) e^{st + ikx}, 
\end{align}
where $k$ is the wave-number and $s$ is the growth rate. 
Using this ansatz, equations \eqref{RBeq:L2} and \eqref{RBeq:L1} become
\begin{align}
\Bigg[ \left( 1  + \mu_2 \frac{\sin^2 2 \theta^{(0)}}{4} + \mu_3 \right)  \left( D^2 - k^2 \right)^2     + \mu_2 \Bigg( i \, \frac{ \sin 4 \theta^{(0)}}{2} &\left( k D^3 - k^3 D \right) \nonumber \\ - \cos 4 \theta^{(0)} k^2 D^2 \Bigg)  \Bigg] w' - \mathcal{R} k^2 T' & =  \frac{s}{\mathcal{P}} \left( D^2 - k^2 \right) w' ,  \label{RBeq:L4a} \\
w' + \left[ D^2 - k^2 \right] T'& = s T',
\label{RBeq:L4b}
\end{align}
where we have adopted the convention $D= \dif / \dif z$.
Equations \eqref{RBeq:L4a} and \eqref{RBeq:L4b} form an eigenvalue problem which must be solved subject to the boundary conditions \eqref{RBeq:ND_BC1} and \eqref{RBeq:ND_BC2} rewritten as
\begin{align}
\begin{split}
w' = D w' &= T' = 0 \text{ on a rigid surface}, \\
w' = D^2 w' &= T' = 0 \text{ on a free surface}.
\end{split} \label{RBeq:L5}
\end{align}

The growth rate $s$ represents an eigenvalue to equations \eqref{RBeq:L4a} and \eqref{RBeq:L4b}, \emph{i.e.}\ for a given dimensionless wave-number $k$ there will be non-trivial solutions ($w',T'$) to equations \eqref{RBeq:L4a} and \eqref{RBeq:L4b} only for certain values of $s$.
We establish for each wave-number $k$ the maximum Rayleigh number $\mathcal{R}_l(k)$ such that the real part of all eigenvalues $s$ are negative, \emph{i.e.}\ the largest Rayleigh number such that the perturbation is stable and any disturbance decays to zero.
The minimum of $\mathcal{R}_l(k)$ is of particular interest, and is termed the critical Rayleigh number ($\mathcal{R}_c$); it is used to determine the physical conditions under which instability first occurs \citep{acheson1990elementary,Drazin,koschmieder1993benard}. If for a given experimental setup $\mathcal{R}<\mathcal{R}_c$ then any perturbation decays exponentially to zero.
The corresponding value of $k$ at $\mathcal{R}_c$ is also of interest; it describes the inverse wave-length of the convection currents and is termed the critical wave-number $(k_c)$.

\section{Numerical solution method}\label{RBsec:Numeric}
In order to determine the marginal stability curves $\mathcal{R}_l(k)$ we must solve the eigenvalue problem \eqref{RBeq:L4a} and \eqref{RBeq:L4b} with boundary conditions given by \eqref{RBeq:L5}. This is achieved using Chebyshev collocation, a spectral method that is capable of achieving high accuracy for low computational cost \citep{trefethen2000spectral}.

Using the Chebyshev differentiation matrix $\bm{D}$ \citep{trefethen2000spectral} the linear operators on $w'$ and $T'$ in equations \eqref{RBeq:L4a} and \eqref{RBeq:L4b} may be approximated. This allows us to form the generalised matrix eigenvalue problem
\begin{align}
\bm{A} \, \bm{x} &= s \, \bm{B} \, \bm{x}, \label{RBeq:GEVP}
\end{align}
where $s$ is the growth rate and eigenvalue of the problem, $\bm{A}$ and $\bm{B}$ are matrices which are discrete representations of the linear operators which act on $w'$ and $T'$, and the vector $\bm{x}$ contains the coefficients of the Lagrange polynomials which approximate $w'$ and $T'$ at the Chebyshev points (equivalently the values of $w'$ and $T'$ at the Chebyshev points) \citep{trefethen2000spectral}.
The matrices $\bm{A}$ and $\bm{B}$ may be constructed in MATLAB for each tuple of parameters $\theta^{(0)}$, $\mu_2$ and $\mu_3$; however, the matrices are not full rank as boundary conditions must be applied to close the problem. These constraints are applied using the method described by \citet{Hoepffner}; the solution space is reduced to consider only interpolants which satisfy the boundary conditions.
We may therefore compute the eigenvalue $s$ for a range of parameters ($\theta^{(0)},\mu_2,\mu_3$) and Rayleigh number $\mathcal{R}$ using the inbuilt eigenvalue solver in MATLAB {\it eig}; this solver employs the $QZ$-algorithm for generalized eigenvalue problems. We then determine the Rayleigh number for which the eigenvalue is zero using the MATLAB function {\it fzero}, \emph{i.e.}\ disturbances neither grow nor decay, and {\it fminsearch} to determine the critical wave and Rayleigh numbers.

As the Prandtl number ($\mathcal{P}$) only appears in combination with the growth rate $s$, we do not consider variations in $\mathcal{P}$ as we are interested in the marginal stability curves where $s=0$, \emph{i.e.}\ the boundary between stability and instability.

To accommodate uncertainty in parameter values, we have performed an extensive parameter search for a wide range of steady state preferred directions ($0 \leqslant \theta^{(0)} \leqslant \pi / 2$) and viscosities ($0 \leqslant \mu_2, \mu_3 \leqslant 1000$). 
Notice that the solution is periodic in the steady state preferred direction with period $\pi/2$.

To validate our numerical procedure we compared our results with those of \citet{dominguez1984marginal} and \citet{rayleigh1916lix} for the Newtonian case, \emph{i.e.}\ $\mu_2=\mu_3=0$; we will denote the critical Rayleigh number for the Newtonian case $\mathcal{R}_N$.
When both boundaries are free, our numerical approximation of the Rayleigh number is within $10^{-12}$ of the known analytical result $\mathcal{R}_N=27 \pi^4/4$. When both boundaries are rigid our numerical approximation of the Rayleigh number is within $10^{-7}$ of the value of $\mathcal{R}_N$ found by \citet{dominguez1984marginal}. 

\section{Results}\label{RBsec:Results}
In section \ref{5.1} we first determine the marginal stability curves $\mathcal{R}_l(k)$; for any value of $k$, an experimental set-up satisfying $\mathcal{R}<\mathcal{R}_l(k)$ is stable for that wavelength, whereas if $\mathcal{R}$ lies above $\mathcal{R}_l(k)$ the system is unstable. We calculate these curves for a range of non-dimensional parameters representing the steady state preferred direction $\theta^{(0)}$, the anisotropic extensional viscosity $\mu_2$ and the anisotropic shear viscosity $\mu_3$ for different combinations of boundary conditions. We determine the critical wave and Rayleigh numbers for each tuple of non-dimensional parameters ($\theta^{(0)}, \mu_2$ and $\mu_3$) by finding the wave-number at which $\mathcal{R}_l(k)$ is minimal. Provided that the Rayleigh number for a given experiment lies below this critical value, the system will be stable to small perturbations for all wavelengths and the fluid will be motionless. In section \ref{5.2} we will make an empirical approximation to the dependence of the critical wave and Rayleigh numbers on the problem parameters.

\subsection{Critical wave and Rayleigh number}\label{5.1}

\begin{figure}
\centering
\includegraphics[width=.8\textwidth]{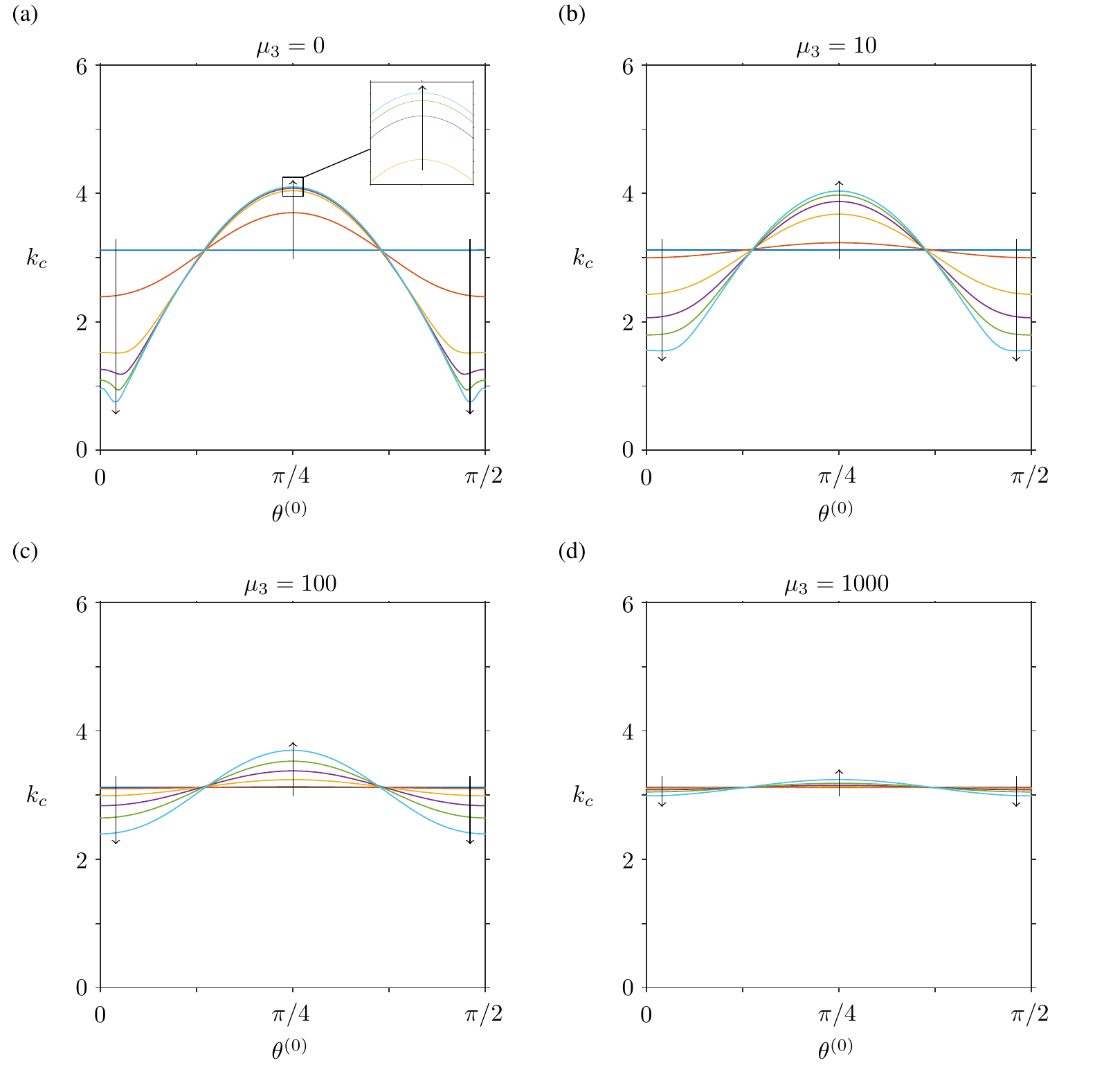}
\caption{Critical wave-number ($k_c$) for changes in the anisotropic extensional viscosity ($\mu_2$), the anisotropic shear viscosity ($\mu_3$), and the preferred direction in the fluid at steady state ($\theta^{(0)}$) with both boundaries rigid. In each subfigure the arrows indicate increasing $\mu_2$ ($\mu_2=0,10,100,250,500,1000$) for (a) $\mu_3=0$, (b) $\mu_3=10$, (c) $\mu_3=100$, and (d) $\mu_3=1000$.}\label{RBfig:rigid_rigid_k}
\end{figure}

\begin{figure}[h]
\centering
\includegraphics[width=.8\textwidth]{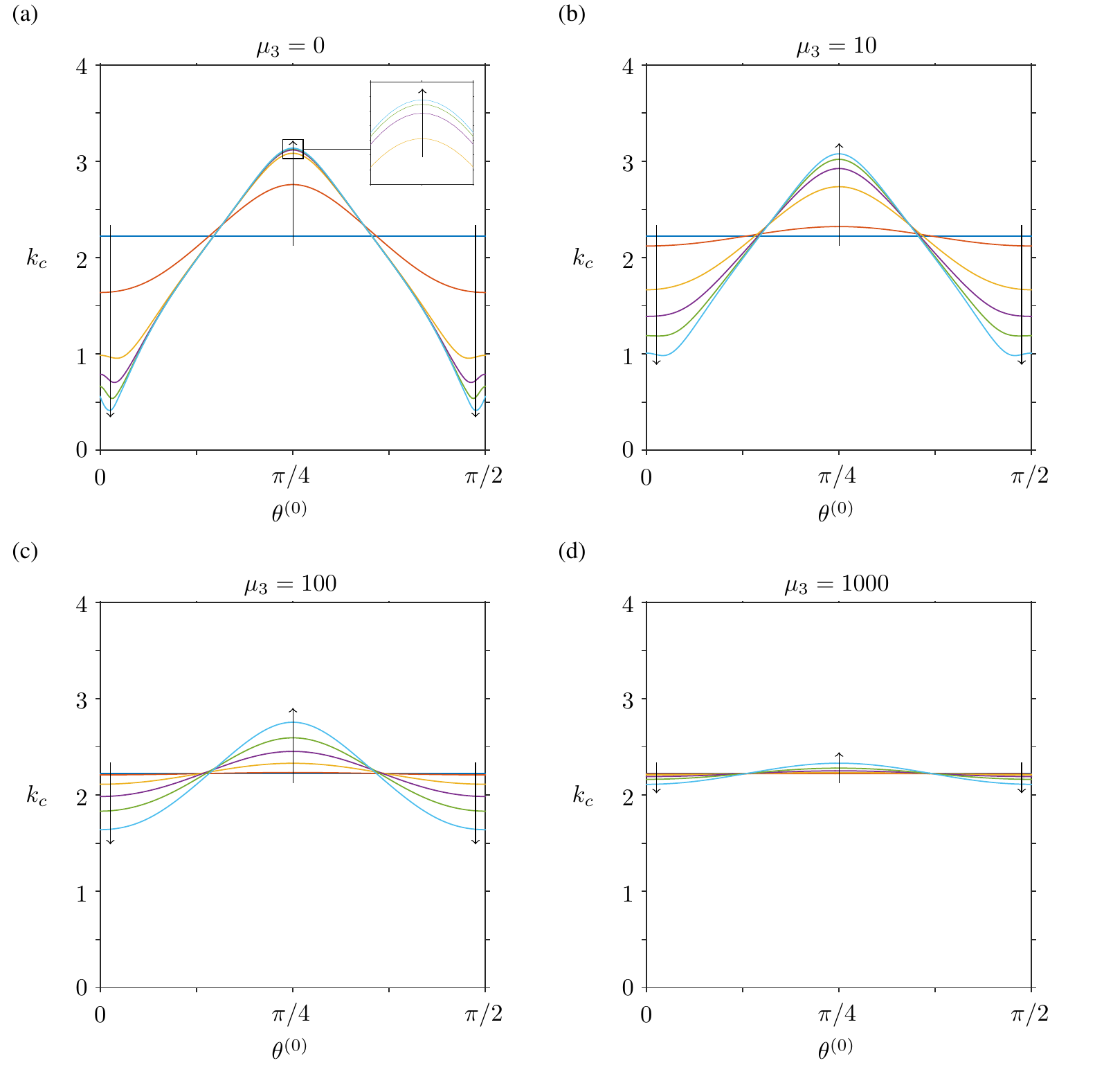}
\caption{Critical wave-number ($k_c$) for changes in the anisotropic extensional viscosity ($\mu_2$), the anisotropic shear viscosity ($\mu_3$), and the preferred direction in the fluid at steady state ($\theta^{(0)}$) with both boundaries free. In each subfigure the arrows indicate increasing $\mu_2$ ($\mu_2=0,10,100,250,500,1000$) for (a) $\mu_3=0$, (b) $\mu_3=10$, (c) $\mu_3=100$, and (d) $\mu_3=1000$.}\label{RBfig:free_free_k}
\end{figure}

Figure \ref{RBfig:rigid_rigid_k} shows the critical wave-number ($k_c$) as a function of the steady state preferred direction ($\theta^{(0)}$), for selected values of the anisotropic extensional ($\mu_2$) and shear ($\mu_3$) viscosities, with both boundaries rigid. The critical wave-number is related to the width of a convection cell; increases in $k_c$ reduce the width of the convection cell. Notice that Figures \ref{RBfig:rigid_rigid_k}(a) -- (d) are symmetric about $\theta^{(0)}=\pi/4$, where the maximum of $k_c$ is achieved. In Figure  \ref{RBfig:rigid_rigid_k}(a) we examine the effect of the anisotropic extensional viscosity with the anisotropic shear viscosity set to zero. The horizontal line corresponds to the Newtonian/isotropic case, and hence there is no dependence on the fibre direction $\theta^{(0)}$. As $\mu_2$ is increased, the limiting form of the critical curve between $\pi/8 \leqslant \theta^{(0)} \leqslant 3\pi/8$ is quickly approached, with changes to $\mu_2$ above $100$ having only a small effect. In the ranges $0\leqslant \theta^{(0)} \leqslant \pi/8$ and $3 \pi/8\leqslant \theta^{(0)} \leqslant \pi/2$, the changes to the critical wave-number occur much more slowly with respect to $\mu_2$, with a local minimum occurring for values of $\mu_2$ above $250$ around $\theta^{(0)}=0.1$ and $\theta^{(0)}=1.5$. The impact of changing the anisotropic extensional viscosity on the wave-number is therefore dependent on the steady state fibre direction. If the fibres are aligned near horizontal or vertical, the wave-number is decreased and the width of the convection cell increased; if the fibre direction is at $\pi/4$ to the horizontal at steady state, then the wave-number increases and hence the width of the convection cell decreases. 
Observing how the critical curves change between Figures \ref{RBfig:rigid_rigid_k}(a) -- (d) allows us to identify the impact of the anisotropic shear viscosity $\mu_3$. As $\mu_3$ is increased it dampens changes to the critical wave-number caused by changes in $\mu_2$, nearly removing the dependence on $\theta^{(0)}$ completely in Figure \ref{RBfig:rigid_rigid_k}(d) where $\mu_3=1000$. 
Similar results are obtained when both boundaries are free (Figure \ref{RBfig:free_free_k}), but where the critical wave-number of a Newtonian fluid ($k_N$) is smaller.

\begin{figure}[h]
\centering
\includegraphics[width=.8\textwidth]{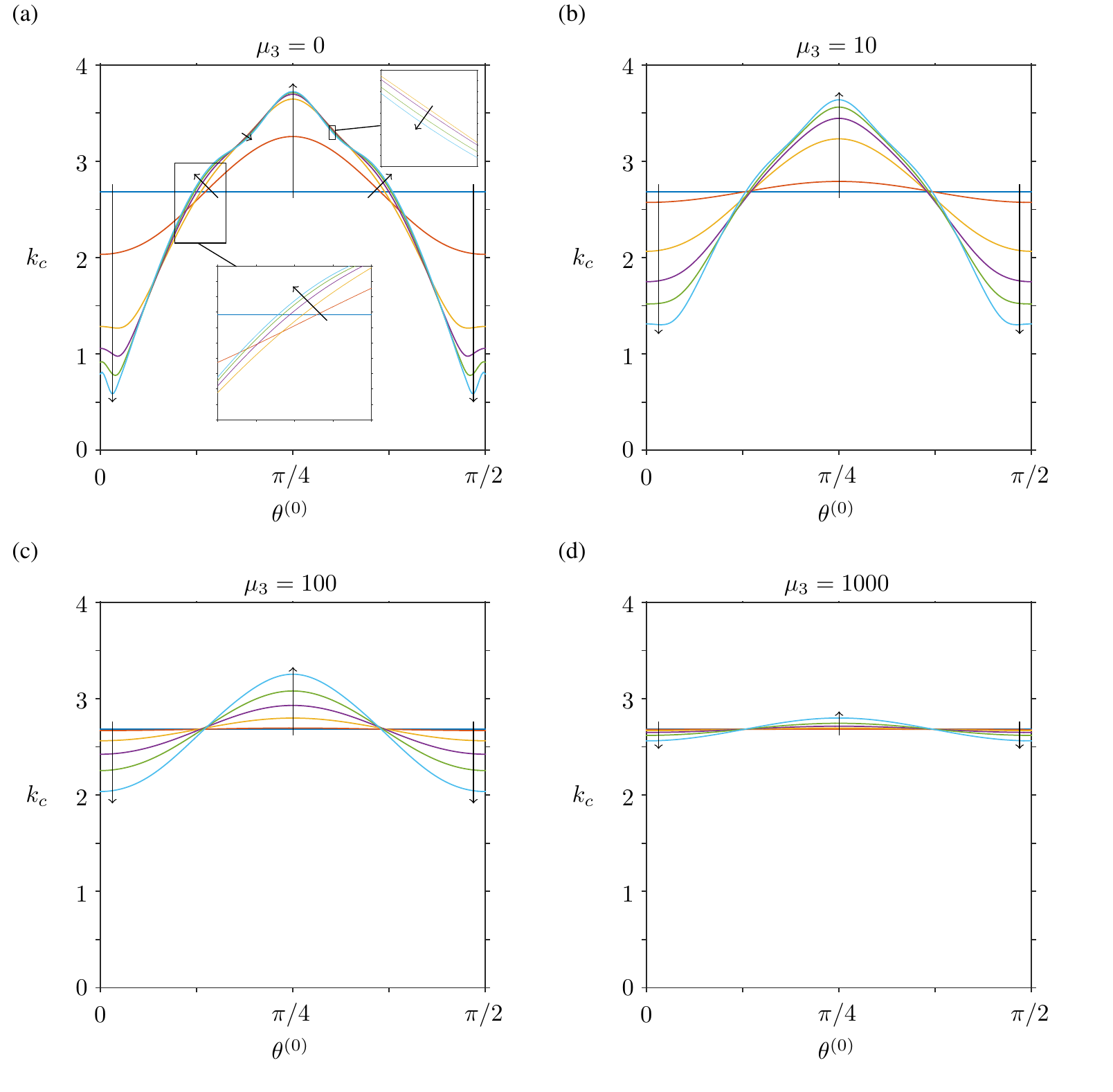}
\caption{Critical wave-number ($k_c$) for changes in the anisotropic extensional viscosity ($\mu_2$), the anisotropic shear viscosity ($\mu_3$), and the preferred direction in the fluid at steady state ($\theta^{(0)}$) with the bottom boundary rigid and the top boundary free. In each subfigure the arrows indicate increasing $\mu_2$ ($\mu_2=0,10,100,250,500,1000$) for (a) $\mu_3=0$, (b) $\mu_3=10$, (c) $\mu_3=100$, and (d) $\mu_3=1000$.}\label{RBfig:free_rigid_k}
\end{figure}

Figure \ref{RBfig:free_rigid_k} shows $k_c$ as a function of $\theta^{(0)}$ for selected values of $\mu_2$ and $\mu_3$ when the lower boundary is rigid and the top free, and shows a more intricate dependence on the tuple of parameters $(\theta^{(0)}, \mu_2, \mu_3)$ than when upper and lower boundaries match. In Figure \ref{RBfig:free_rigid_k}(a) the horizontal line corresponds to the Newtonian/isotropic case, and hence has no dependence upon $\theta^{(0)}$, as expected. As $\mu_2$ is increased the critical curves become more complex, in the range $0 \leqslant \theta^{(0)} \leqslant \pi/8$ and $3 \pi/8 \leqslant \theta^{(0)} \leqslant \pi/2$ similar behaviour is observed to when both boundaries are the same, with the appearance of a local maximum at $\theta^{(0)}=0,\pi/2$ and a global minimum for values of $\theta^{(0)}=0.1,1.4$. However, for $\theta^{(0)}$ between $\pi/8$ and $\pi/4$ an extra mode is introduced compared with the matching boundary cases, but this variation becomes small for values of $\mu_2$ larger than $100$. 
We again identify from Figures \ref{RBfig:free_rigid_k}(a) -- (d) that $\mu_3$ dampens the change in the critical wave-number due to $\mu_2$, eventually removing the dependence on $\theta^{(0)}$ (Figure \ref{RBfig:free_rigid_k}(d)).

\begin{figure}[h]
\centering
\includegraphics[width=.8\textwidth]{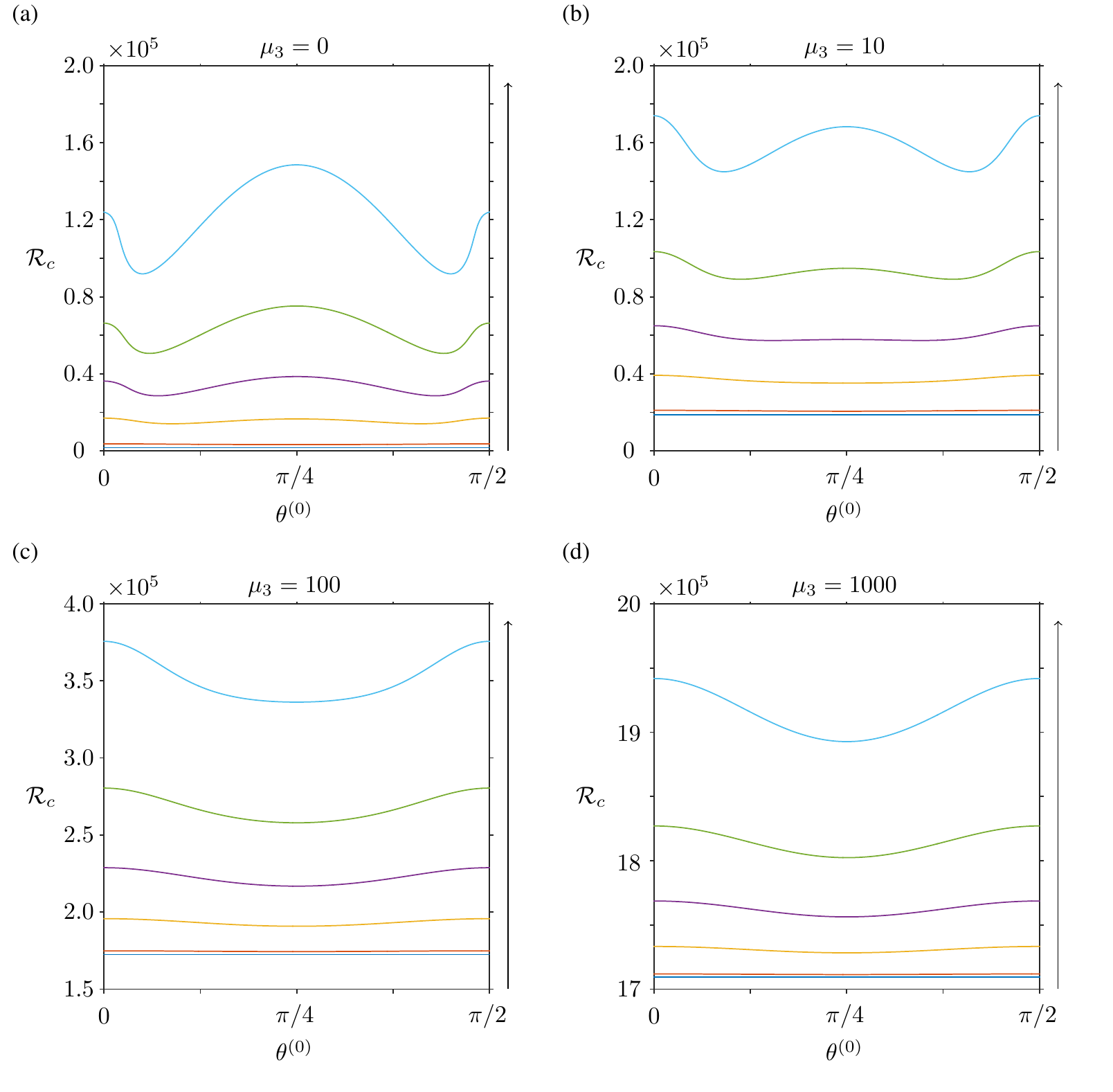}
\caption{Critical Rayleigh number ($\mathcal{R}_c$) for changes in the anisotropic extensional viscosity ($\mu_2$), the anisotropic shear viscosity ($\mu_3$), and the preferred direction in the fluid at steady state ($\theta^{(0)}$) with both boundaries rigid. In each subfigure the arrows indicate increasing $\mu_2$ ($\mu_2=0,10,100,250,500,1000$) for (a) $\mu_3=0$, (b) $\mu_3=10$, (c) $\mu_3=100$, and (d) $\mu_3=1000$.}\label{RBfig:rigid_rigid_R}
\end{figure}

Figure \ref{RBfig:rigid_rigid_R} shows the critical Rayleigh number ($\mathcal{R}_c$) as a function of $\theta^{(0)}$ for changes in $\mu_2$ and $\mu_3$ wtih both boundaries rigid. Figure \ref{RBfig:rigid_rigid_R}(a) shows the change in $\mathcal{R}_c$ neglecting anisotropic shear viscosity, \emph{i.e.}\ $\mu_3=0$.
The lowest horizontal line corresponds to the Newtonian case, and has no dependence on $\theta^{(0)}$ as expected. For $\mu_2 \lesssim 100$ this horizontal line is simply translated to higher values of $\mathcal{R}_c$, with little to no dependence on $\theta^{(0)}$. 
As $\mu_2$ is increased further the shape of the critical curves change dramatically. Global minima occur at $\theta^{(0)} \approx 0.3, 1.2$, local maxima occur at $\theta^{(0)}=0, \pi/2$, and the global maximum at $\theta^{(0)}=\pi/4$; the difference between the global minimum and maximum is approximately $6\times10^4$ for $\mu_2=1000$. Therefore when the anisotropic extensional viscosity is large and the anisotropic shear viscosity is negligible the steady state is most unstable for steady state fibre orientations close to $\pi/16$ of horizontal or vertical, however for smaller angles to the horizontal or vertical the stability sharply increases. The most stable case when the steady state direction is at $\pi/4$ to the horizontal.
Examining Figures \ref{RBfig:rigid_rigid_R}(a) -- (d) allows us identify how the anisotropic shear viscosity affects the stability of the steady state. 
We observe that increasing $\mu_3$ increases $\mathcal{R}_c$, hence making the steady state more stable. However, this relationship is not uniform for different values of $\theta^{(0)}$, as can be seen by noting that when $\mu_2=1000$ and $\mu_3=0$ the most stable value of $\theta^{(0)}$ is $\pi/4$, but as $\mu_3$ is increased to $1000$ then $\theta^{(0)}=\pi/4$ becomes the most unstable value. 
Therefore increasing the anisotropic shear viscosity has the most stabilising effect for values of the steady state fibre orientation near horizontal and vertical, and a slightly weaker effect when the steady state direction is $\pi/4$. However, increases in the anisotropic shear viscosity always stabilise the steady state for all choices of anisotropic extensional viscosity and steady state preferred directions.

\begin{figure}[h]
\centering
\includegraphics[width=.8\textwidth]{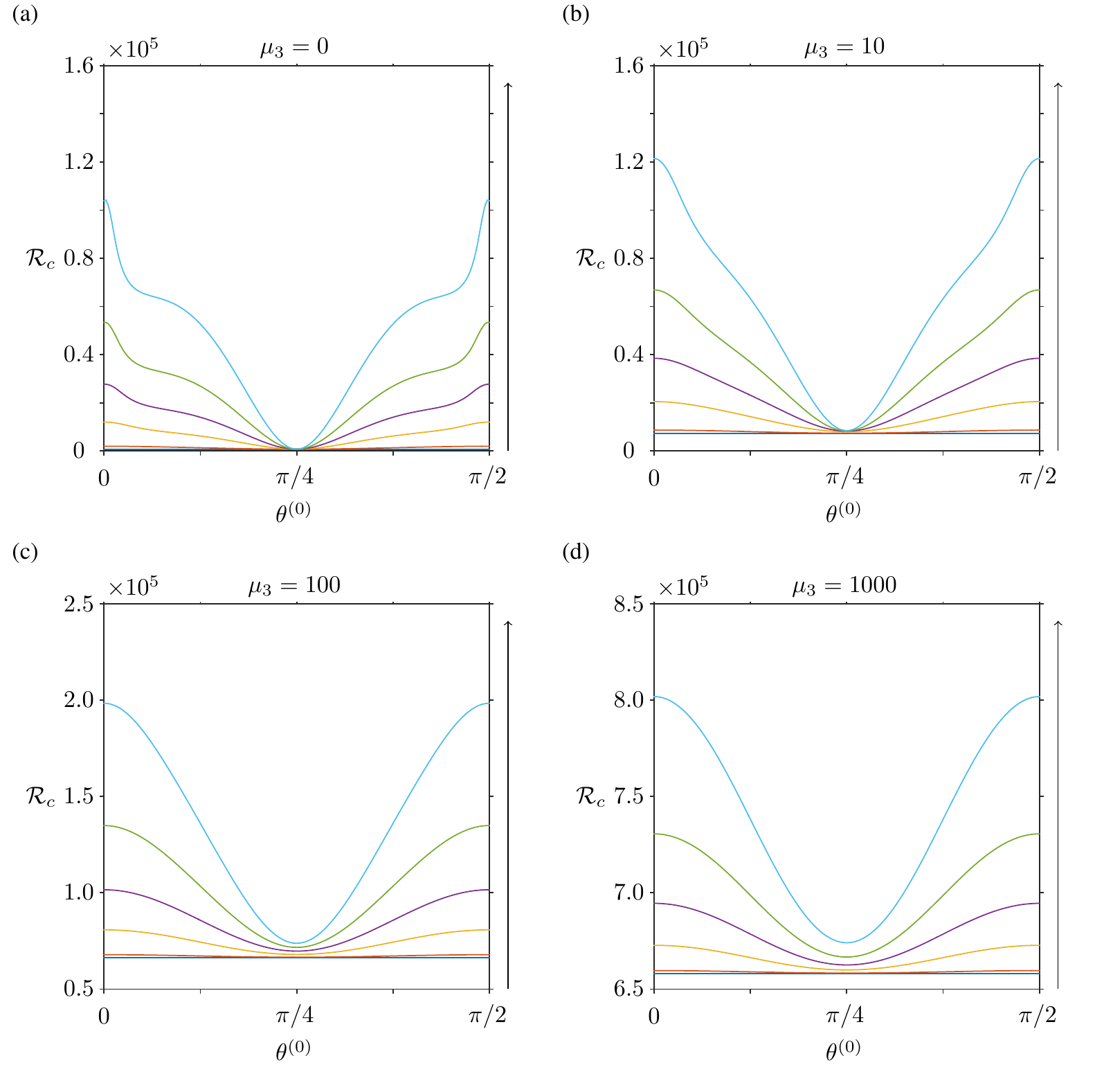}
\caption{Critical Rayleigh number ($\mathcal{R}_c$) for changes in the anisotropic extensional viscosity ($\mu_2$), the anisotropic shear viscosity ($\mu_3$), and the preferred direction in the fluid at steady state ($\theta^{(0)}$) with both boundaries free. In each subfigure the arrows indicate increasing $\mu_2$ ($\mu_2=0,10,100,250,500,1000$) for (a) $\mu_3=0$, (b) $\mu_3=10$, (c) $\mu_3=100$, and (d) $\mu_3=1000$.}\label{RBfig:free_free_R}
\end{figure}

Figure \ref{RBfig:free_free_R} shows the dependence of $\mathcal{R}_c$ on $\theta^{(0)}$ for selected values of $\mu_2$ and $\mu_3$ with both boundaries free.
In Figure \ref{RBfig:free_free_R}(a) $\mu_3=0$ and the Newtonian case is represented by the lowest horizontal line. 
As $\mu_2$ is increased a global maximum occurs at $\theta^{(0)}=0$ and $\pi/2$ and global minimum at $\theta^{(0)}=\pi/4$, where $\mathcal{R}_c$ does not increase from the critical Rayleigh number for the Newtonian/isotropic case ($\mathcal{R}_c \approx \mathcal{R}_N$). Near horizontal or vertical fibre-orientation, increasing the anisotropic extensional viscosity increases the threshold at which instability occurs, but when the steady state preferred direction is $\pi/4$ there is little change to the stability threshold as anisotropic extensional viscosity is varied.
Figures \ref{RBfig:free_free_R}(a) -- (d) show that as $\mu_3$ is increased, $\mathcal{R}_c$ increases regularly, smoothing out the points of inflection that occur for small values of $\mu_2$. Therefore, increasing $\mu_3$ stabilises the steady state for all values of $\theta^{(0)}$ and $\mu_2$. Changes in anisotropic shear viscosity affect the magnitude of the critical Rayleigh number much more than changes to the anisotropic extensional viscosity.

\begin{figure}[h]
\centering
\includegraphics[width=.8\textwidth]{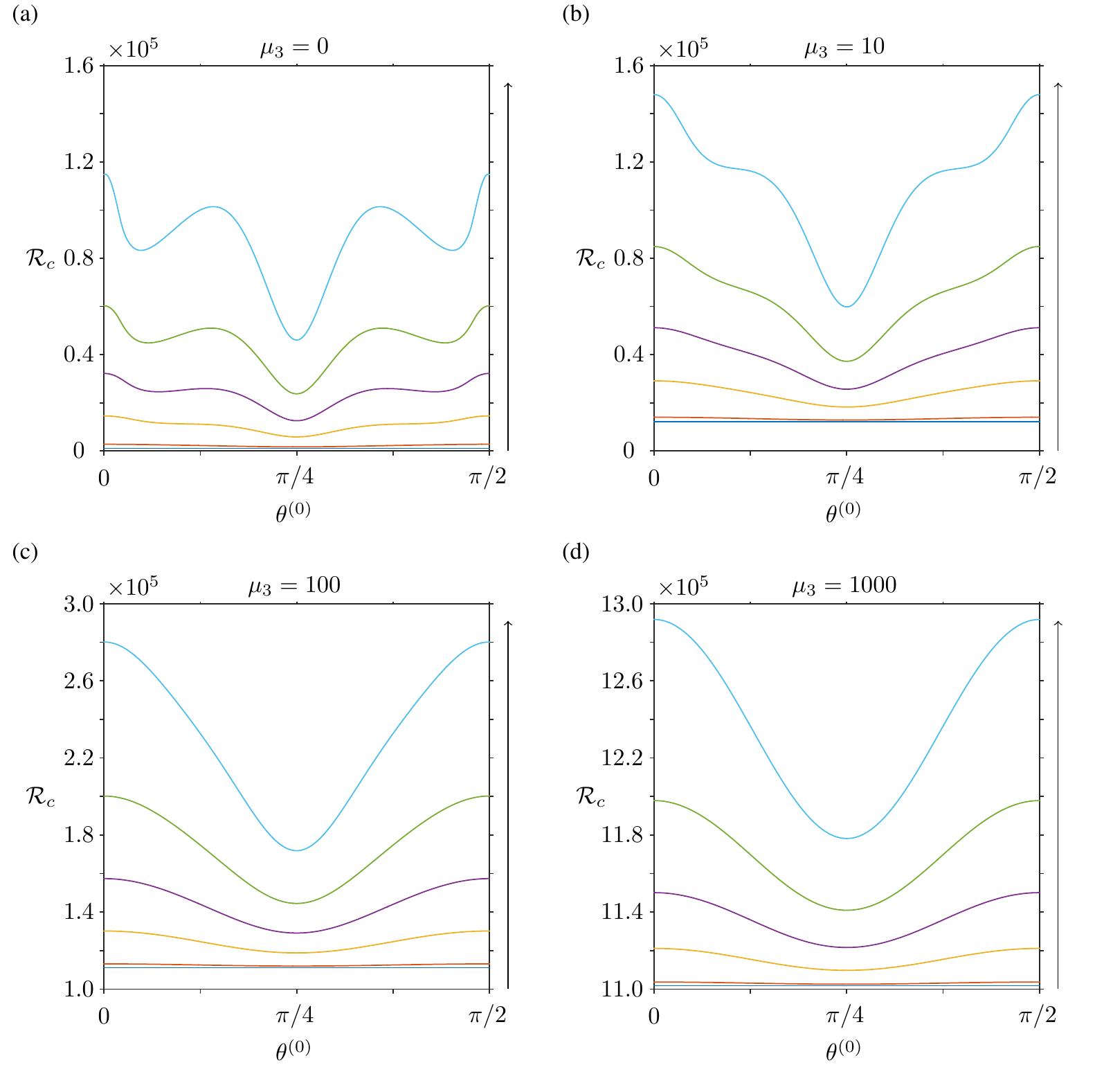}
\caption{Critical Rayleigh number ($\mathcal{R}_c$) for changes in the anisotropic extensional viscosity ($\mu_2$), the anisotropic shear viscosity ($\mu_3$), and the preferred direction in the fluid at steady state ($\theta^{(0)}$) with the bottom boundary rigid and the top boundary free. In each subfigure the arrows indicate increasing $\mu_2$ ($\mu_2=0,10,100,250,500,1000$) for (a) $\mu_3=0$, (b) $\mu_3=10$, (c) $\mu_3=100$, and (d) $\mu_3=1000$.}\label{RBfig:free_rigid_R}
\end{figure}

Figure \ref{RBfig:free_rigid_R} shows the dependence of $\mathcal{R}_c$ on $\theta^{(0)}$ for selected values of $\mu_2$ and $\mu_3$, with the lower boundary rigid and the upper free. 
In Figure \ref{RBfig:free_rigid_R}(a) we examine how $\mu_2$ affects $\mathcal{R}_c$ when $\mu_3=0$; the Newtonian/isotropic case is shown by the lowest horizontal line. As $\mu_2$ is increased, $\mathcal{R}_c$ increases however this increase is not uniform with respect to $\theta^{(0)}$. Global maxima occur at $\theta^{(0)}=0$ and $\pi/2$, local maxima at $\theta^{(0)} \approx \pi/8$ and $ 3 \pi / 8$, local minima at $\theta^{(0)} \approx \pi/16$ and $ 7 \pi / 16$, and the global minimum at $\theta^{(0)}=\pi/4$. Therefore increasing the anisotropic extensional viscosity increases the stability threshold most when, at steady state, the fibres are either horizontal or vertical, and least when they are directed at an angle $\pi/4$ radians.
As $\mu_3$ increases, $\mathcal{R}_c$ is increased, with the additional local maxima and minima becoming less pronounced, and disappearing completely once $\mu_3 \gtrsim 100$, as can be identified by comparing Figures \ref{RBfig:free_rigid_R}(a) -- (d). Again, changes in $\mu_3$ affect $\mathcal{R}_c$ far more than similar changes to $\mu_2$. Therefore increases in the anisotropic shear viscosity stabilise the steady state, with the most stabilisation occurring when the fibres are oriented horizontally or vertically at steady state, at least least when $\theta^{(0)} = \pi/4$.

Comparing Figures \ref{RBfig:rigid_rigid_k} -- \ref{RBfig:free_rigid_R} allows us to examine the effect of the boundary conditions on the critical wave and critical Rayleigh numbers. 
Examining Figures \ref{RBfig:rigid_rigid_k} and \ref{RBfig:free_free_k} we identify that when the top and bottom boundaries are the same, the curves for the critical wave-number take the same form, but with lower critical wave-numbers for the free-free boundaries than the rigid-rigid case. When the boundaries are mixed, and the anisotropic shear viscosity is negligible, two additional modes occur between $\pi/8 \leqslant \theta^{(0)} \leqslant 3 \pi/8$. However, variation between the critical curves is small for medium to large values of the anisotropic extensional viscosity, and all changes are dampened as the anisotropic shear viscosity is increased, similarly to the matching boundary case. Therefore, in all cases, the anisotropic extensional viscosity gives rise to variations in the critical wave-number with respect to the steady state preferred direction, which are dampened by increases in the anisotropic shear viscosity.

Comparing Figures \ref{RBfig:rigid_rigid_R} -- \ref{RBfig:free_rigid_R} allows us to compare how the different boundary conditions affect the critical Rayleigh number. Similarly to the Newtonian/isotropic case the most stable pair of boundaries is rigid-rigid, with the most unstable being free-free. In all boundary pairs increasing either the anisotropic extensional or shear viscosities increases the critical Rayleigh number, however changes to the anisotropic shear viscosity affect the stability threshold much more than equivalent changes to the anisotropic extensional viscosity. 

We notice in Figures \ref{RBfig:rigid_rigid_R} -- \ref{RBfig:free_rigid_R} that the critical wave and Rayleigh numbers are the same for $\theta^{(0)}$ and $\pi/2 - \theta^{(0)}$, \emph{i.e.}\ the material has the same stability characteristics when the steady state preferred direction is horizontal or vertical, the dependance on this angle being symmetric about $\pi/4$. When the transversely-isotropic material is interpreted as a suspension of elongated particles, this result may be explained by noting that when the particles are either horizontal or vertical, only translational, non-rotating, motion will be induced. The stability characteristics of these states should therefore be similar. 

\subsection{Empirical forms of critical curves}\label{5.2}
Examining Figures \ref{RBfig:rigid_rigid_k} -- \ref{RBfig:free_rigid_R} we notice that, for medium to large values of the anisotropic shear viscosity, the critical curves are continuous with no sharp extrema (\emph{i.e.}\ we expect the rate of change of the critical values with $\theta^{(0)}$ to be continuous also). We may therefore attempt to fit analytic functions to the numerical results, for $k_c$ and $\mathcal{R}_c$. The critical wave-number is fit by minimising the maximum absolute error between the function and the numerical results through the simplex search method of \citet{lagarias1998convergence} ({\it fminsearch} in MATLAB); the critical Rayleigh number $\mathcal{R}_c$ via the trust region method of nonlinear least squares fitting ({\it fit} in MATLAB).

\begin{figure}
\centering
\includegraphics[width=.8\textwidth]{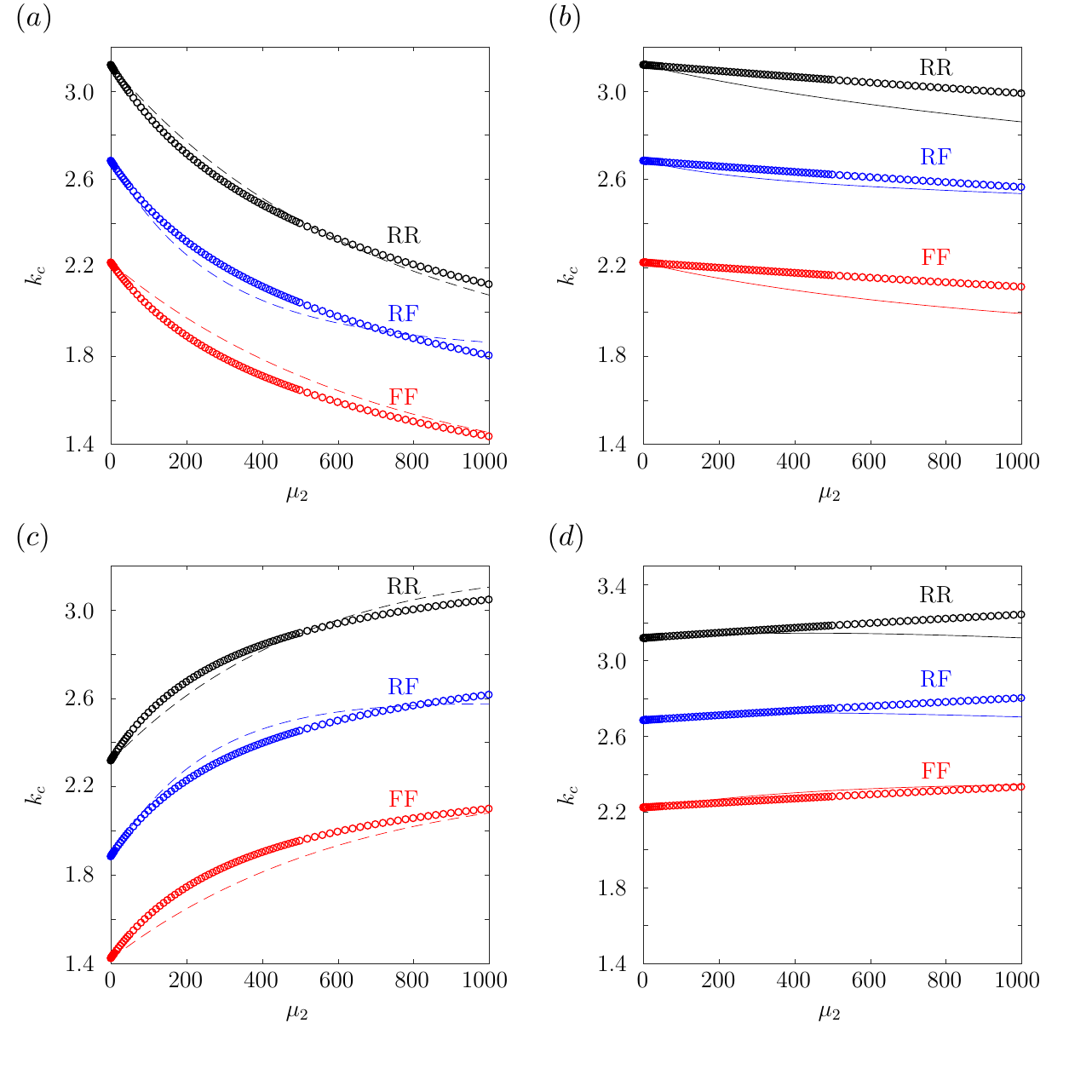}
\caption{Comparison of fitted curves with the numerical results for the critical wave-number $k_c$. The fit for (a) ($\mu_3=50$, $\theta^{(0)}=0$) and (b) ($\mu_3=1000$, $\theta^{(0)}=0$), (c) ($\mu_3=50$, $\theta^{(0)}=\pi/4$), (d) ($\mu_3=1000$, $\theta^{(0)}=\pi/4$) where black, blue and red correspond to rigid-rigid ($RR$), rigid-free ($RF$) and free-free ($FF$) boundary pairs and circles represent numerical results. We choose $\theta^{(0)}=0,$ $\pi/4$ so that we consider the maximum and minimum values of $\cos 4 \theta^{(0)}$.} \label{figureKcFullCurves}
\end{figure}

For $\mu_3>40$ we fit the critical wave-number to the empirical form 
\begin{align}
k_c (\mu_2,\mu_3, \theta^{(0)})\approx \frac{f_1 \left( \exp \left(-f_5 \mu_2 \right) -1 \right)}{f_2+\mu_3}\cos(4 \theta^{(0)})-\frac{f_3 \mu_2}{f_4+\mu_3}+k_N,\label{RBeq:k_fit}
\end{align}
where $k_N$ is the critical wave-number for a Newtonian fluid and $f_1$ to $f_5$ are fitting parameters.
The form of this function implies that for $\mu_3 \gg \mu_2$, and $\mu_3 \gg 1$, the critical wave-number approximates its Newtonian value. For $\mu_2=0$ the critical wave-number $k_c$ does not depend on $\theta^{(0)}$ or $\mu_3$, and takes the same value as in the Newtonian case. Figure \ref{figureKcFullCurves} shows a comparison between the sampled numerical results and fitted function, where the fitted parameters are given in table \ref{tableFitk}; excellent qualitative and good quantitative agreement is found.

\begin{table}
\centering
\caption{The parameter values from curve fitting for the critical wave-number given in equation \eqref{RBeq:k_fit} for the different combinations of boundary conditions.}\label{tableFitk}
\begin{tabular}{ |c|c|c|c|c|c| }
\hline
{\bf Boundary Type} & $f_1$ & $f_2$  &  $f_3$ & $f_4$&$f_5$ \\
\hline 
rigid-rigid & $ 175.9$ & $107.3$ &  $12.8 \times 10^{-3}$  & $50.8$  & $1.7 \times 10^{-3}$ \\

rigid-free & $91.1$ & $67.7$ & $5.3 \times 10^{-3}$ & $33.8$ & $3.8 \times 10^{-3}$ \\

free-free & $284.4$ & $259.5$ & $3.7 \times 10^{-3}$ & $19.5$ & $1.5 \times 10^{-3}$ \\
\hline
\end{tabular}

\end{table}

\begin{figure}
\centering
\includegraphics[width=.8\textwidth]{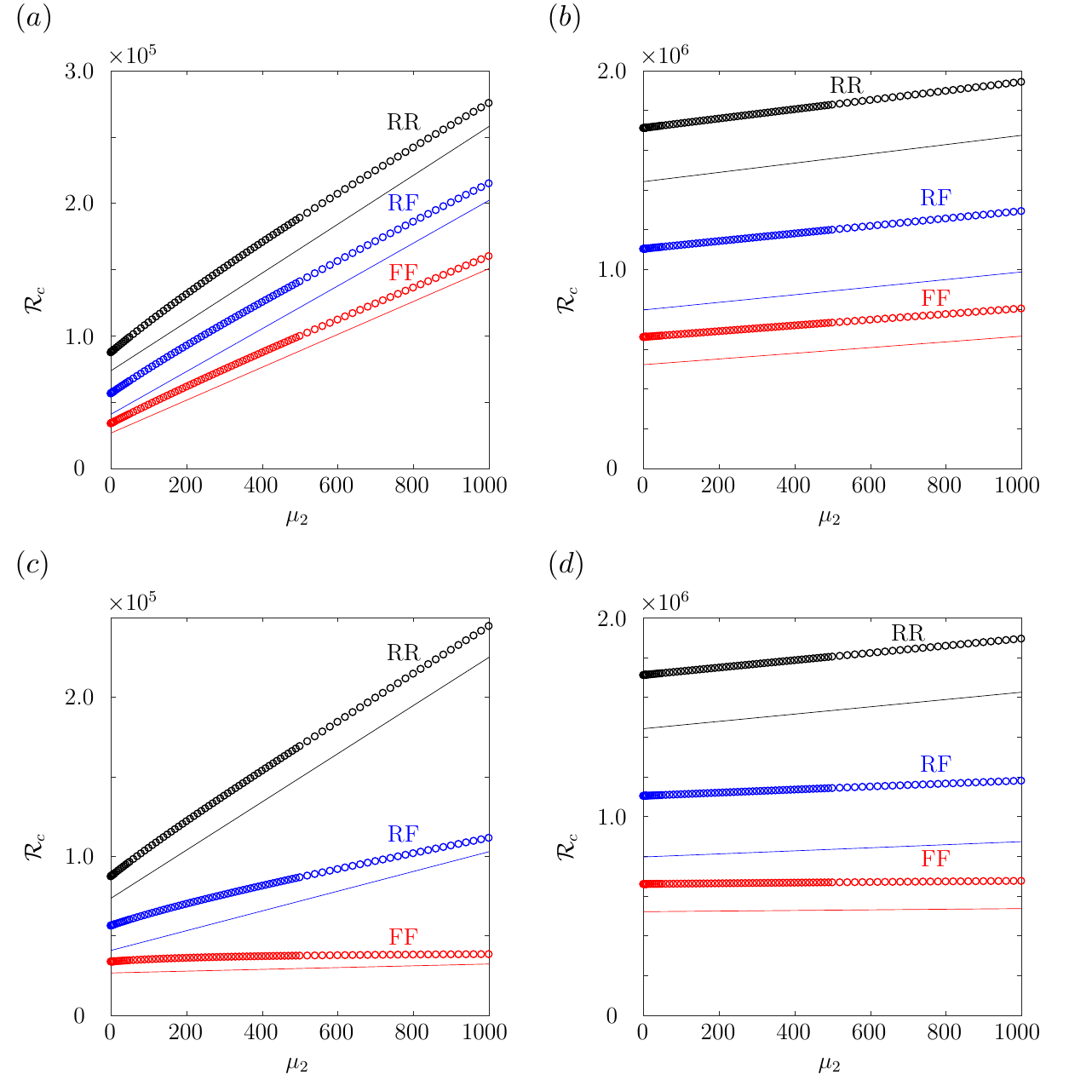}
\caption{Comparison of fitted curves with the numerical results for the critical Rayleigh number $\mathcal{R}_c$. The fit for (a) ($\mu_3=50$, $\theta^{(0)}=0$) and (b) ($\mu_3=1000$, $\theta^{(0)}$) (c) ($\mu_3=50$, $\theta^{(0)}=\pi/4$) (d) ($\mu_3=1000$, $\theta^{(0)}=\pi/4$), where black, blue and red correspond to rigid-rigid ($RR$), rigid-free ($RF$) and free-free ($FF$) boundary pairs and circles represent the numerical results. We choose $\theta^{(0)}=0,$ $\pi/4$ so that we consider the maximum and minimum values of $\cos 4 \theta^{(0)}$.} \label{figureRcFullCurves}
\end{figure}

For $\mu_3>40$ we fit the critical Rayleigh number to the empirical form 
\begin{align}
\mathcal{R}_c (\mu_2, \mu_3, \theta^{(0)}) \approx \left( \frac{-g_1}{\mu_3 + 1} + g_2 \right) \mu_2 \cos (4 \theta^{(0)})+\left( \frac{-g_3}{\mu_3^{1/2} + 1} +g_4 \right)\mu_2+\left( g_5 \mu_3 + g_6 \right), \label{RBeq:R_fit}
\end{align}
where $g_1$ to $g_6$ are fitting parameters.
The form of this function implies that $\mathcal{R}_c$ is dependent upon $\theta^{(0)}$ for $\mu_2 \gg \mu_3$, however this dependence is dampened as $\mu_3$ is increased. We also observe from the relative sizes of $g_4$ and $g_5$, in table \ref{tableRC}, that the increase of $\mathcal{R}_c$ with $\mu_3$ is much greater than that with $\mu_2$. 
Figure \ref{figureRcFullCurves} shows a comparison between the sampled numerical results and fitted function, with the fitted values found in table \ref{tableRC}; a reasonable quantitative agreement and a good qualitative agreement is found. Note that $g_6$ plays the role of $\mathcal{R}_N$, and numerically is extremely close to its known Newtonian values.

\begin{table}
\centering
\caption{The parameter values from curve fitting for the critical Rayleigh number in equation \eqref{RBeq:R_fit} for the different combinations of boundary conditions.}\label{tableRC}
\begin{tabular}{ |c|c|c|c|c|c|c| }
\hline
{\bf Boundary Type} & $g_1$ & $g_2$  &  $g_3$ & $g_4$&$g_5$&$g_6$ \\
\hline 
rigid-rigid & $448.4$ & $25.2$ &  $431.6$  & $221.4$  & $1440.8$& $1706.4$ \\

rigid-free & $388.8$ & $57.3$   & $238.5$ & $141.3$ & $796.0$&$1099.9$ \\

free-free & $247.5$ & $64.1$    & $156.1$ & $84.2$ & $521.1$ &$657.0$\\
\hline
\end{tabular}

\end{table}

\section{Conclusion}\label{RBsec:Discussion}
In this paper, we extended the work of \citeauthor{rayleigh1916lix} to study the linear stability of a transversely-isotropic viscous fluid, contained between two horizontal boundaries, which are either rigid or free, of different temperatures. 
We used the stress tensor first proposed by \citet{ericksen1960transversely}, with $\mu_1=0$ (equivalent to a passive fluid \citep{holloway2017influences}), and a kinematic equation for the fibre-director field to model a transversely-isotropic fluid. Numerically, we presented results for a range of steady state, initially uniform, preferred fibre directions from horizontal to vertical; this is equivalent to the full range of directions as the governing equations have a period of $\pi/2$.

As found recently for the Taylor-Couette flow of a transversely-isotropic fluid \citep{holloway2015linear}, the anisotropic shear viscosity $\mu_3$, is much more important in determining the stability of the flow than the anisotropic extensional viscosity $\mu_2$. The influence of this pair of parameters upon the stability of the flow depends on the uniform steady state preferred direction $\theta^{(0)}$, as well as the boundary conditions. Similarly to a Newtonian fluid, the most stable pair of boundaries is rigid-rigid, for which the temperature difference between the two boundaries required to induce instability is the largest; the least stable boundary pair is free-free.

The rheological parameters $\mu_2$ and $\mu_3$ also have an impact on the critical wave-number $k_c$, which describes the width of the convection cells. We find for steady state preferred directions near horizontal or vertical, the width of the convection cell increases with the anisotropic extensional viscosity, when compared to a Newtonian fluid, and decreases when the preferred direction makes an angle of $\pi/4$ with the horizontal. The anisotropic shear viscosity dampens any changes to the critical wave-number caused by increases in the anisotropic extensional viscosity. If $\mu_3\gg \mu_2$ and $\mu_3 \gg 1$ then there is very little change to the critical wave-number, and hence convection cell size, with changes to the anisotropic extensional viscosity or steady state preferred direction.

We are able to fit empirical functions which exhibited excellent qualitative and good quantitative agreement for critical wave-number and good qualitative and reasonable quantitative agreement for critical Rayleigh number. The relative parameter values emphasised the relative importance of anisotropic extensional and shear viscosities. Empirical functions of this type may be valuable in making predictions regarding fibre-reinforced flows without the need to resort to expensive computation.

The analysis we have undertaken in this paper shows that the stability characteristics of a transversely-isotropic fluid are significantly different from those of a Newtonian fluid. Therefore when the fluid exhibits a preferred direction, such as a fibre-laden fluid, these effects should be taken into account.

Fluids which exhibit transversely-isotropic rheology are commonly found in many industrial and biological applications, therefore it is necessary to gain a better understanding of the underlying mechanics governing the behaviour of these materials. As a classical fluid mechanics problem modified to incorporate anisotropic rheology, we hope the Rayleigh-B\'{e}nard stability analysis undertaken here will motivate research into this fascinating area.

\section*{Acknowledgment}
CRH is supported by an Engineering and Physical Sciences Research Council (EPSRC) doctoral training award (EP/J500367/1) and RJD the support of the EPSRC grant (EP/M00015X/1). The authors thank Gemma Cupples for valuable discussions.

\label{lastpage}

\end{document}